\documentclass[Afour,sagev,times]{sagej}

\usepackage{moreverb,url}
\usepackage{pifont}
\usepackage{algorithm}
\usepackage{algorithmic}
\usepackage{subcaption}
\usepackage{soul}
\usepackage{float}
\usepackage{placeins}
\usepackage{xcolor}
\usepackage[draft,colorlinks,bookmarksopen,bookmarksnumbered,citecolor=red,urlcolor=red]{hyperref}
\usepackage{multirow}
\usepackage{makecell}
\usepackage{rotating}

\newcommand\BibTeX{{\rmfamily B\kern-.05em \textsc{i\kern-.025em b}\kern-.08em
T\kern-.1667em\lower.7ex\hbox{E}\kern-.125emX}}

\def\volumeyear{2016}

\begin{document}
\runninghead{Raza et al.}

\title{EmoWrite: A Sentiment Analysis-Based Thought to Text Conversion - A Validation Study}

\author{Imran Raza\affilnum{1}, Syed Asad Hussain\affilnum{1}, Muhammad Hasan Jamal\affilnum{1},  \\
Isabel de la Torre Diez \affilnum{2}, Carmen Lili Rodriguez Velasco \affilnum{3,4,5}, Jose Manuel Brenosa \affilnum{3,6,7} and Imran Ashraf \affilnum{8,*} }

\affiliation{\affilnum{1}Department of Computer Science, COMSATS University Islamabad, Lahore Campus, Lahore, Pakistan.\\
\affilnum{2}Department of Signal Theory, Communications and Telematics Engineering. Unviersity of Valladolid, Paseo de Belen, 15. 47011 Valladolid – Spain.\\
\affilnum{3}Universidad Europea del Atlantico. Isabel Torres 21, 39011 Santander,
Spain.\\
\affilnum{4}Universidad Internacional Iberoamericana Campeche 24560, Mexico.\\
\affilnum{5}Fundacion Universitaria Internacional de Colombia Bogota, Colombia.\\
\affilnum{6}Universidad Internacional Iberoamericana Arecibo, Puerto Rico 00613, USA.\\
\affilnum{7}Universidade Internacional do Cuanza. Cuito, Bie, Angola. \\
\affilnum{8} Department of Information and Communication Engineering, Yeungnam University, Gyeongsan 38541, South Korea; (imranashraf@ynu.ac.kr)}
\corrauth{Imran Ashraf}
\email{imranashraf@ynu.ac.kr}

\begin{abstract}
\textbf{Objective: }
The objective of this study is to introduce "EmoWrite," a novel brain-computer interface (BCI) system aimed at addressing the limitations of existing BCI-based systems. Specifically, the objective includes improving typing speed, accuracy, user convenience, emotional state capturing, and sentiment analysis within the context of BCI technology.

\textbf{Method:}
The method involves the development and implementation of EmoWrite, utilizing a user-centric Recurrent Neural Network (RNN) for thought-to-text conversion. The system incorporates visual feedback and introduces a dynamic keyboard with a contextually adaptive character appearance. Comprehensive evaluation and comparison against existing approaches are conducted, considering various metrics such as accuracy, typing speed, sentiment analysis, emotional state capturing, and user interface latency. The data required for this experiment was obtained from a total of 72 volunteers (40 male and 32 female) aged between 18 and 40.

\textbf{Results: }
EmoWrite achieves notable results, including a typing speed of 6.6 Words Per Minute (WPM) and 31.9 Characters Per Minute (CPM) with a high accuracy rate of 90.36\%. It excels in capturing emotional states, surpassing other systems with an Information Transfer Rate (ITR) of 87.55 bits/min for commands and 72.52 bits/min for letters. It also offers an intuitive user interface with a low latency of 2.685 seconds.

\textbf{Conclusion:}
The introduction of EmoWrite represents a significant stride towards enhancing BCI usability and emotional integration.  The findings indicate that EmoWrite shows promising potential in enhancing communication methods, with future implications for individuals with motor disabilities.
\end{abstract}


\maketitle

\section{Introduction}
Mind-reading systems were fiction, but with the advancement of technology, are becoming a reality and helping physically challenged people in performing their tasks e.g., controlling a wheelchair, robotic arm, and cursor. People having a physical disability and speech obstruction are not disabled, rather they are differently abled because they might be impaired with one or two abilities, but their other capabilities can be more precise and accurate as compared to a healthy person \cite{1}. Particularly, when talking about paralytic patients, might not have an appropriate communication medium to convey their feelings, but their mental activity is more precise if it is utilized efficiently. A lot of research work has been done which is not limited only to assisting in rehabilitation, but to making them self-reliant \cite{2}. Scientists are utilizing brain signals, usually through electroencephalogram (EEG), by extracting useful information from acquired signals. These brain signals are being used in many domains of daily life and exclusively in the medical domain for monitoring alertness, coma or death, brain damage, controlling anesthesia depth, brain development, testing of drugs, and monitoring sleep disorders \cite{2}. EEG signals are also used to resolve speech impediments and eradicate communication barriers of paralytic patients by converting their thoughts (silent speech) to text. 

There are two methods used in literature to decode brain signals. The first method directly decodes brain signals into a word, while the second method requires the use of an intermediate output device for converting thought to text. Converting a word directly from the brain to text seems not so feasible because only limited numbers of words can be interpreted at a time due to the need for additional training, computation power, and resources. Limited information is available regarding the aspect that whether the brain generates the same signals while perceiving similar words or not. Hence, this research area has not yet matured. \cite{3} in their research decode five words only i.e. “Alpha”, “Bravo”, “Charlie”, “Delta” and “Echo” while \cite{4} only decode five characters i.e. a, e, i, o, t. The second method needs a medium, which includes an interface containing characters or words, that can be selected with the help of brain signals. The character selection can be based on two mechanisms i.e. using a virtual keyboard or Visual Evoked Potential (VEP)/Steady-State Visual Evoked Potential (SSVEP) \cite{5}. The virtual keyboard uses raw data or built-in functions of the headset i.e. left, right, up, down, or motor imagery (i.e. imaging movements of hands or feet) \cite{2}, whereas attention-based systems (VEP/SSVEP) focus on some flickering stimulus for selection of characters \cite{5}. The factors that can affect the performance of these systems are related to speed, accuracy, and usability of the system. Moreover, people using these systems are not able to express their feelings accurately because it is problematic to find a proper word to write according to one’s mood, so incorporation of the emotional state along with other commands from the brain will help in better utilization of these systems.

Using a deep learning approach can be significantly important in this regard and make potential contributions. For example, the study \cite{uppal2023enhancing} introduced a technique to predict brain strokes with high accuracy. The brain stroke data was used to build the model which is a multilayer perceptron (MLP). It employed multiple optimizers, including adaptive moment estimation with maximum, root mean squared propagation (RMSProp), and the adaptive learning rate method. Experimental results indicate that the MLP model combined with the RMSProp optimizer performed the best, achieving a training accuracy of 95.8\% and a testing accuracy of 94.9\%. Another possible venture is the self-supervised learning phenomenon where large-sized unlabeled datasets are used for model training. The existing research is lacking on self-supervised learning, the study \cite{abdulrazzaq2024consequential} discusses applications of self-supervised learning from the prospect of industrial engineering and medicine. The authors identify the key possibilities for prediction in these fields using self-supervised learning. Analysis is carried out in the context of medical staff predicting patient ailments more efficiently, without relying on traditional numerical models that require a lot of computation, time, storage, and effort for data annotation. Similarly, functional near-infrared spectroscopy (fNIRS) has been used for brain-computer interface (BCI) tasks. The study \cite{zafar2023metaheuristic} introduces a method for selecting important features for brain-computer interface (BCI) applications using functional near-infrared spectroscopy (fNIRS). Temporal statistical features, like the mean, slope, maximum, skewness, and kurtosis, were calculated from all channels to create a training vector. Seven different optimization algorithms were tested for their ability to classify data using a k-nearest neighbor cost function: particle swarm optimization, cuckoo search optimization, the firefly algorithm, the bat algorithm, flower pollination optimization, whale optimization, and grey wolf optimization (GWO). This method was tested on an online dataset of motor imagery (MI) and mental arithmetic (MA) tasks from 29 healthy subjects. The results showed that using the features selected by these optimization algorithms significantly improved classification accuracy compared to using all available features.

The proposed innovative system "EmoWrite" seamlessly integrates a dynamic and personalized graphical user interface with the capability to predict contextually relevant words based on the individual's mood. The pivotal innovation of EmoWrite lies in its ability to monitor the emotional states of users and facilitate the articulation of their emotions through words. Due to nuanced differentiations among emotional classes, precise modeling becomes imperative. These emotional classes exhibit variations from person to person, thereby necessitating the successful resolution of the significant challenge of person-specific emotional class detection. The classification of brain signals, pivotal for identifying emotional classes, demands meticulous training. The adaptive arrangement of characters on the keyboard is designed to streamline character selection, facilitating swift typing. Furthermore, the character set arrangement adapts to the user's unique typing style and contextual cues, enhancing the efficiency of communication for differently abled patients. For signal acquisition, the Emotiv Epoc+ headset, equipped with 14 EEG sensing channels, is employed to capture brain signals. EmoWrite implements established classification techniques sourced from existing literature, optimizing training efficiency. Given the deterioration of facial expressions over time in paralytic patients due to decreased or absent usage, the proposed system also encompasses emotion detection alongside the utilization of facial expressions. This dual functionality bears potential benefits for the rehabilitation process. The contributions of this paper are as follows:
\begin{itemize}
    \item Introduces BCI-driven solutions with the potential to support individuals with severe disabilities, focusing on initial validation and future applications for those with paralysis and speech impairments.
    \item Translates inner speech into text using a dynamic keyboard featuring context-adaptive character displays.
    \item Presents an innovative character arrangement on the keyboard, streamlining character selection and enhancing typing speed for users.
    \item Enables sentiment-guided thought-to-text conversion and recommendations, a novel feature not previously documented in existing literature.
\end{itemize}

The rest of the paper is organized as follows. Section \ref{LR} describes the related work. Section \ref{MTH} discusses the proposed scheme for data acquisition and data processing. Section \ref{ER} describes 
 the real-time experimentation and results followed by the conclusion in Section \ref{con} that reveals the potential of EmoWrite to convert silent speech to text.
\vspace{-2mm}
\section{Related Work}
\label{LR}
To enable communication for paralytic patients some work has been done in the past and still research is ongoing in the domain of BCI-based thoughts-to-text conversion. One of the methods to convert thoughts into text is using a graphical user interface (GUI) consisting of numbers, alphabet, or special characters, which are displayed in a certain order on a virtual keyboard. 

\newpage
\renewcommand{\arraystretch}{1.2}
\begin{sidewaystable*}[htbp]
\scriptsize
\centering
\caption{Comparison of Related Work}
\begin{tabular}{|c|c|c|c|c|c|c|c|l|c|c|c|c|c|}\hline
 & \multicolumn{2}{c}{Virtual Keyboards} & \multicolumn{4}{|c}{Attention Based} & \multicolumn{2}{|c}{Keyboard Layout} & \multicolumn{2}{|c|}{Action Selection} &  &   & \\ \cline{2-11}
Ref.	&	Flickering & 	Simple 	& SSVEP & VEP & Eye Gaze & \makecell[c]{Attention \\ Level} & Static & Dynamic	& \makecell[c]{Raw \\ Data}	&	\makecell[c]{Built-in \\ Functions}	& \makecell[c]{Emotional \\ State} & Accuracy & CPM/WPM \\ \hline 
Zhang et al. (2018) \cite{2}  &   -	      &	\ding{52}	&	-	    & \ding{52} &	-	   &	-	   & \ding{52} &	-	   &	-	   & \ding{52} &	-	   & 95.53\% & 6.67 CPM \\ \hline
Gupta et al. (2019) \cite{8}  &	\ding{52} &    -      	& \ding{52} & \ding{52} &	-	   & 	-	   &    -      &	-	   & \ding{52} &	-	   & \ding{52} & 74.95\% &   N/A    \\ \hline
Masud et al. (2017) \cite{16} &	\ding{52} &    -      	&     -     & \ding{52} &	-	   & 	-	   & \ding{52} &	-	   &	-	   &	-	   & 	-	   & 87.50\% &   N/A    \\ \hline
Chen et al. (2015) \cite{17} &	\ding{52} &    -      	& \ding{52} &	-	    &	-	   & 	-	   &	-	   &	-	   &	-	   &	-	   & 	-	   &   N/A   &  12 WPM  \\ \hline
Cecotti (2011) \cite{5}  &	\ding{52} &    -      	& \ding{52} & \ding{52} &	-	   & 	-	   &	-	   &	-	   & \ding{52} &	-	   & 	-	   &   N/A   &  \makecell[c]{5-10 CPM(P300) \\ 7.34 CPM (SSVEP) \\ 5 CPM (motor)}  \\ \hline
Chen et al. (2014) \cite{18} &	\ding{52} &    -      	& \ding{52} &	-	    &	-	   & 	-	   & \ding{52} &	-	   &	-	   &	-	   & 	-	   & 80-90\% &  6.5 WPM  \\ \hline
Sp{\"u}ler et al. (2012) \cite{19} &	\ding{52} &    -      	&   -      	& \ding{52} &	-	   & 	-	   & \ding{52} &	-	   &	-	   &	-	   & 	-	   & 96\% &  9 WPM  \\ \hline
Higger et al. (2016) \cite{20} &	\ding{52} &    -      	&   -      	&	-	    &	-	   & 	-	   & \ding{52} &	-	   &	-	   &	-	   & 	-	   & 94\%  &  N/A  \\ \hline
Akce et al. (2014) \cite{21} &	\ding{52} &    -      	&   -      	& \ding{52} &	-	   & 	-	   & 	-	   & \ding{52} &	-	   &	-	   & 	-	   & N/A &  11.93 CPM  \\ \hline
Cecotti (2016) \cite{22} &    -      	  & \ding{52}   &   -      	&   -      	&\ding{52} & 	-	   & 	-	   & 	-	   &	-	   &	-	   & 	-	   & N/A &  9.3 CPM  \\ \hline
Ben-Ami et al. (2019) \cite{23} &    -      	  &  -          &   -      	&   -      	&	-	   &\ding{52}  & 	-	   & 	-	   &\ding{52}  &	-	   & 	-	   & 25\% &  N/A  \\ \hline
Alomari et al. (2014) \cite{6}  &    -      	  & \ding{52}   &   -      	&   -      	&	-	   &	-	   & \ding{52} & 	-	   &\ding{52}  &	-	   & 	-	   & N/A &  N/A  \\ \hline
Hayet et al. (2019) \cite{7}  &    -      	  & \ding{52}   &   -      	&   -      	&	-	   &	-	   & 	-	   & 	-	   &\ding{52}  &	-	   & 	-	   & N/A &  N/A  \\ \hline
Williamson et al. (2009) \cite{9}  &    -      	  & \ding{52}   &   -      	&   -      	&	-	   &	-	   & 	-	   & 	-	   & 	-	   & \ding{52} & 	-	   & N/A &  7 CPM  \\ \hline
Wang et al. (2018) \cite{4}  &    -      	  &  -          &   -      	&   -      	&	-	   &	-	   & 	-	   & 	-	   &\ding{52}  &	-	   & 	-	   & 31\% &  N/A  \\ \hline
Jarosiewicz et al. (2015) \cite{24}  &    -      	  & \ding{52}   &   -      	&   -      	&	-	   &	-	   & 	-	   & 	-	   & 	-	   & \ding{52} & 	-	   & N/A &  12 CPM  \\ \hline
Pandarinath et al. (2017) \cite{10}  &    -      	  & \ding{52}   &   -      	&   -      	&	-	   &	-	   & 	-	   & 	-	   & \ding{52} & 	-	   & 	-	   & N/A &  \makecell[c]{36 CPM (QWERTY) \\ 39 CPM (OPTI-II) \\ 13.5 CPM (alphabetic)}  \\ \hline
Arijit et al. (2013) \cite{25}  &    -      	  & \ding{52}   &   -      	&   -      	&	-	   &	-	   & 	-	   & 	-	   & \ding{52} & 	-	   & 	-	   & N/A &  N/A  \\ \hline
Topal et al. (2012) \cite{11}  &    -      	  & \ding{52}   &   -      	&   -      	&	-	   &	-	   & 	-	   & 	-	   & \ding{52} & 	-	   & 	-	   & N/A &  N/A  \\ \hline
Pathirana et al. (2018) \cite{12}  &    -      	  & \ding{52}   &   -      	&   -      	&	-	   &	-	   & 	-	   & 	-	   & 	-	   & \ding{52} & 	-	   & N/A &  6.61 CPM  \\ \hline
Andi et al. (2018) \cite{13}  &    -      	  &    -        &   -      	&   -      	&	-	   &	-	   & 	-	   & 	-	   & \ding{52} & 	-	   & 	-	   & 59.20\% &  N/A  \\ \hline
Birbaumer et al. (2000) \cite{14}  &    -      	  & \ding{52}   &   -      	&   -      	&	-	   &	-	   & 	-	   & 	-	   & \ding{52} & 	-	   & 	-	   & N/A &  N/A  \\ \hline
George et al. (2014) \cite{26}  &    -      	  &    -      	&   -      	&   -      	&	-	   &	-	   & 	-	   & 	-	   & 	-	   & \ding{52} & 	-	   & N/A &  N/A  \\ \hline
Mackenzie et al. (2010) \cite{15} &	\ding{52} &    -      	&   -      	&   -      	&	-	   & 	-	   & 	-	   &   -      	&	-	   &	-	   & 	-	   & 99\% &  5.11 WPM  \\ \hline
Morooka et al. (2018) \cite{27}  &    -      	  &    -      	&   -      	&   -      	&	-	   &	-	   & 	-	   & 	-	   & \ding{52} & 	-	   & 	-	   & 79.90\% &  N/A  \\ \hline
EmoWrite   &    -      	  & \ding{52}   &   -      	&   -      	&	-	   &	-	   & 	-	   & 	-	   & \ding{52} & \ding{52} & \ding{52} & 90.36\% &  \makecell[c]{6.58 WPM \\ 31.92 CPM}  \\ \hline
\end{tabular}
\label{summary}
\end{sidewaystable*}

\newpage

The brain signals are then used to control the selection of any desired character or alphabet from this virtual keyboard. The major selection methods used in previous systems can be categorized as 1) attention-based control like Visual Evoked Potential (VEP) or Steady-State Visual Evoked Potential (SSVEP), and 2) raw data or built-in functions of the headset to control the cursor or targeted area on the screen.
The virtual keyboards are divided into single or multiple layers with static or dynamic keys and their design has a direct influence on the performance of the system. A wide literature survey has been conducted to specify different types of action selection methodologies; character arrangement and virtual keyboard designs.

Some of the major challenges for decoding EEG signals are low signal-to-noise ratio, time consumption, and accuracy. To overcome these challenges, a novel hybrid deep learning approach based on Recurrent Neural Networks (RNNs) and Convolutional Neural Networks (CNNs) is used that converts thoughts to text \cite{2}. EEG signals are used to control the cursor of a personal computer \cite{6}. Features from EEG signals are extracted with a 64-channel NeuroSky headset using a discrete wavelet transform and are classified using machine learning algorithms such as Support Vector Machine (SVM) and Neural Networks (NNs). \cite{7} describe a two-layer hierarchical layout of the keyboard that works with motor imagery signals (left-hand raised, right-hand raised, nodding up and nodding down).

\cite{8} use EEG signals to enhance a written sentence with detected emotions by inserting words in it. Long Short-Term Memory (LSTM) Networks-based language modeling framework is used to verify the sentence correctness by ranking the generated suggestions. ‘Hex-o-spell’, a tilt-based gestural text entry system is introduced by \cite{9} where the letters are arranged in 6 hexagonal shaped boxes that are rearranged after every transition to save time. \cite{4} show how English alphabets are decoded using EEG phase information by using the 64-channel actiCHamp Brain Product to acquire EEG signals. Five alphabets (a, e, l, o, t) are chosen, and against each alphabet, the brain signals are recorded. Results showed that accuracy increased to 31\% and the time improved to 200ms. Results prove that most decrypted data lies within the period of 100 to 600ms. High-performance intracortical BCI for communication is described by \cite{10} which provides point-and-click control of the computer. These controls are translated by the ReFIT Kalman Filter (RKF) which translates the 2D cursor movement and the Hidden Markov Model (HMM) which translates the selection.

Efficiency issues of the virtual keyboard are discussed by \cite{11}. The authors suggest that there should be at least one level of hierarchy for better usability while higher efficiency is achieved with a matrix-shaped keyboard. A novel virtual keyboard design is introduced by \cite{12} using built-in functions of Emotiv Insight that help navigate through the interface and move the selected area. The characters are arranged circularly to utilize screen space efficiently. The dynamic caption of keys changes according to the previously entered characters using a predictive system.

\begin{figure*}[ht!]
\centering
\includegraphics[width=1.0\textwidth]{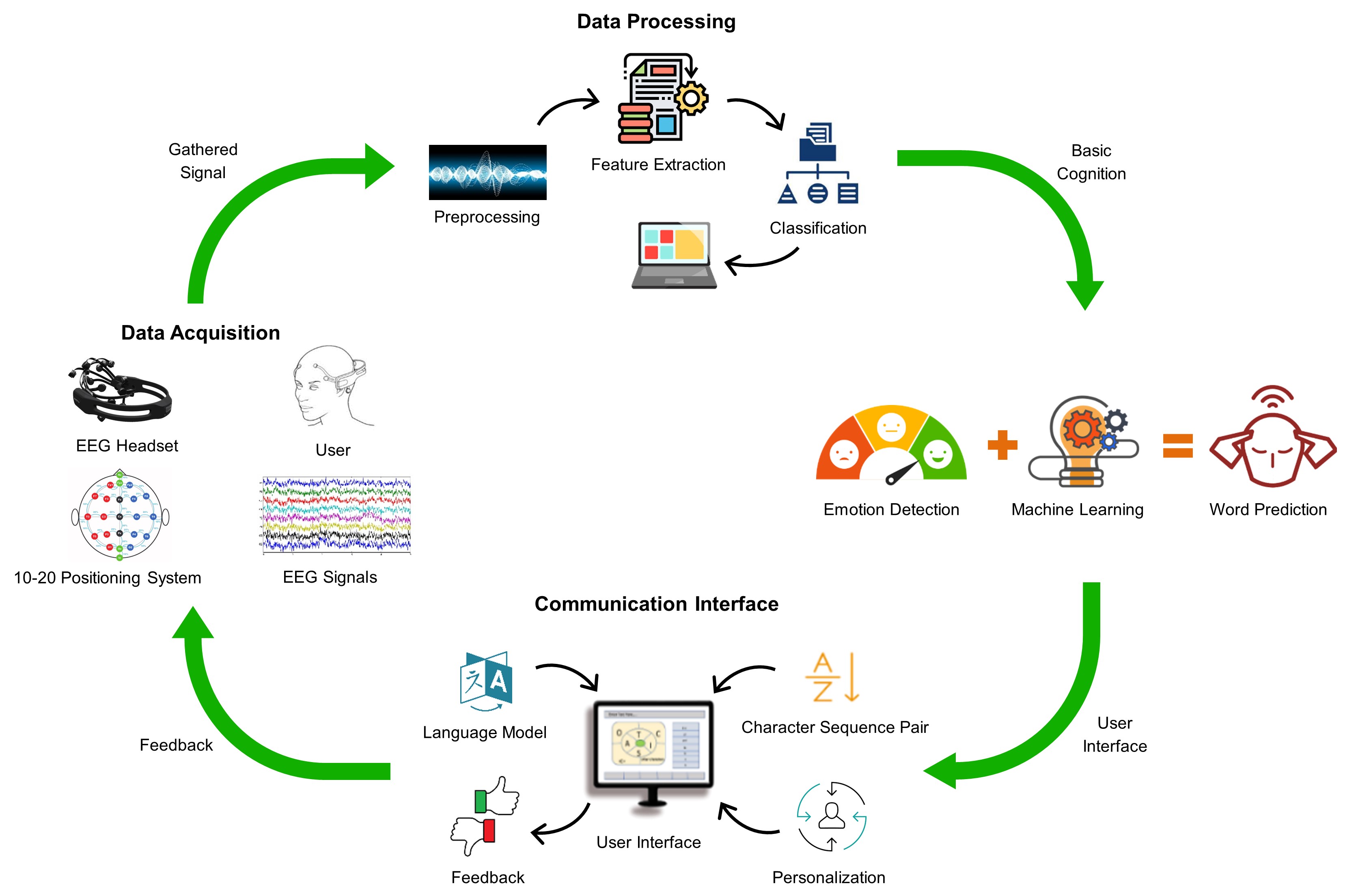}
\caption{Basic Flow of Proposed System}
\label{flowchart}
\end{figure*}

Disabled people have limited activities and need a certain medium to translate their brain signals to interact with people around them. \cite{13} use raw EEG data with Emotiv Epoc to translate thoughts to the text which will be implemented in SMS to provide ease of communication. A Thought Translation Device (TTD) that uses Slow Cortical Potential (SCP) to select characters or words is introduced by \cite{14}. SCP is used because its learning rules are well-known, and the basics are well-understood. Cognitive performance drops when the user produces positive SCP while improvement in the performance and learning occur with negative SCP. To record brain activity an 8-channel EEG amplifier is used. Visual feedback from EEG is received and updated every 63 ms. \cite{15} presents a scanning ambiguous keyboard that takes input from the user through one key or switch. The layout contains a letter section at the top and a word section at the bottom (candidate list). The focus is transferred between the letter section to the word section with the space key. The alphabets are highlighted in a sequence and the user can select them by triggering the input. 

Table \ref{summary} shows the major parameters to check the performance of thought-to-text-based systems. Most existing schemes have implemented simple keyboards rather than visually evoked or flashing characters to achieve high accuracy. Moreover, the traversing of a keyboard can be easily controlled by using simple built-in functions or raw data, instead of utilizing attention-based systems with flashing or flickering characters. That is because attention-based systems require users to dwell on the desired character for a certain amount of time and it also adds training overhead. Furthermore, these systems have not incorporated the emotional state of the patients, which can be integrated to provide efficient and personalized thoughts to the text conversion system.

\section{Methods}
\label{MTH}
Communication medium plays an important role in human-to-human (H2H) or human-to-machine (H2M) interaction. BCI aids paralytic patients, who cannot communicate, by providing solutions for H2H and H2M interaction. Existing BCI-based work in this domain, especially thought-to-text conversion, is limited in efficiency, accuracy, and number of words per minute. Till now the maximum of 12 WPM has been achieved with a non-invasive technique \cite{17}.

Considering all the discussed challenges, EmoWrite integrates a dynamic keyboard with a circular arrangement of keys. The traversal in the proposed keyboard is controlled by mapping brain commands with facial expressions and using the built-in functions of the headset. It also predicts the next helping verb which is displayed on the right side of the screen. Moreover, the next word prediction is emotion-based as well as personalized. Emotion-based predictions of words assist paralytic patients by efficiently converting their thoughts to text. Furthermore, the machine learning algorithm keeps on retraining itself after a specific interval to predict only the latest and up-to-date words. Additionally, integrating the emotional states of patients with machine learning techniques enhances the performance and productivity of the system.

Implementation of EmoWrite aims at reducing the typing delay, increasing accuracy, typing speed, and convenience of the interface. The signals from the brain are acquired through EEG and decoded to convert thoughts to text after extracting information by processing the signals. The extracted information from the brain signals is then classified and mapped to the mental commands (e.g., thinking of left or right direction) or facial expressions (e.g., eye blink, frown, etc.) to perform specific tasks. Emotion state detection has been integrated with machine learning for better productivity of the system. The personalized dynamic arrangement of characters on the screen uses a language model (character sequence pair) and a machine-learning algorithm to show only the desired characters on the screen. Finally, the user gets visual feedback through the typed text shown on the screen and the machine learning algorithm also gets feedback from the user to help update its weights for future predictions. EmoWrite is comprised of the following four modules: 1) Data Acquisition 2) Data Processing 3) Basic Cognition, and 4) Communication Interface. The basic flow of the proposed system is shown in Figure \ref{flowchart}.

\subsection{Data Acquisition}
The primary step in any BCI application is to gather data in the form of brain signals. The process of signal acquisition is performed with different techniques as discussed earlier. In this study, the non-invasive technique is employed, which is riskless and easy to handle. It includes collecting brain signals from the surface of the scalp with the help of an EEG headset. Different versions of dry and wet electrodes-based headsets including Emotiv Insight, Emotiv Epoc+, NeuroSky, MindWave, etc. are available. This study deploys a wet electrodes-based 14-channel Emotiv Epoc+ headset.

The data for this experiment was collected from 72 volunteers (55.6\% male and 44.4\% female), with a mean age of 29 years and a standard deviation of 6.5 years, at the Advanced Communication Networks Lab at COMSATS University Islamabad, Lahore Campus. None of the participants had any previous BCI experience. Before the data collection process, informed consent was obtained from all participants and the study protocol was approved and supervised by the Ethics Committee of COMSATS University Islamabad, Lahore Campus. All experiments were performed under relevant guidelines and regulations. The duration of the study was two years from 2019 to 2021.

\subsection{Data Processing}
The acquired brain signals comprise data on multiple mental activities, but EmoWrite focuses only on data regarding emotional states, mental commands, and facial expressions. To extract meaningful information, the data acquired through the EEG headset is processed using the built-in pre-processing and classification techniques of Emotiv Applications like Emotiv BCI \footnote{Emotiv BCI.” https://www.emotiv.com/emotiv-bci/}, Emotiv PRO \footnote{https://www.emotiv.com/emotivpro/}, and EmotivBrainViz \footnote{https://www.emotiv.com/emotiv-brainviz/}.

\begin{figure*}[ht!]
    \centering
    \begin{subfigure}[b]{4.4cm}
    \includegraphics[width=4.4cm]{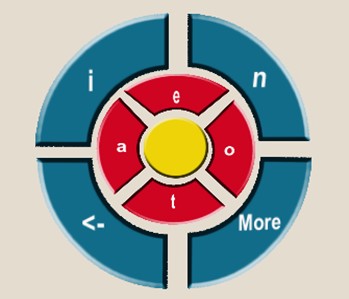}
	\caption{First Set of Characters.}	
    \label{fig4a}
    \end{subfigure}
    \hspace{40pt}
	\begin{subfigure}[b]{4.4cm}	
	\includegraphics[width=4.4cm]{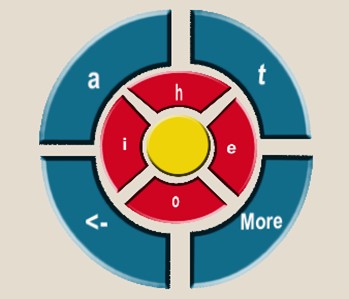}
	\caption{Character set after selection of 't'.}
    \label{fig4b}
    \end{subfigure}
\caption{Keyboard Interface.} 
\label{interface}
\end{figure*}

\subsubsection{Emotion Detection}
The proposed EmoWrite system utilizes the cortex API's performance metrics to detect emotions \footnote{https://emotiv.gitbook.io/emotivpro/data\_streams/performance-metrics} that furnishes crucial insights into a user's cognitive state through the classification of six key metrics; stress, engagement, interest, excitement, focus, and relaxation. These metrics collectively contribute to the identification of four distinct emotional classes namely, happiness, sadness, anger, and calm. Upon ascertaining the detected emotional class, a correlation finder module identifies a collection of emotion-related words from multiple datasets that have been annotated with emotions 
\cite{8}, \cite{28}. To ensure data cleanliness, the dataset undergoes preprocessing, which involves the removal of extraneous spaces and symbols. Next, separate Recurrent Neural Network (RNN) models are trained for each emotional class. These models are designed to facilitate emotion-based predictions. The selection of contextualized words depends on the emotional states inferred, and these chosen words are then integrated into the list of predictive words. This comprehensive methodology leads to an emotion-sensitive word prediction system, enhancing the accuracy and contextuality of the predicted words.


\subsubsection{User-centric Machine Learning Algorithm}
A machine learning algorithm is used along with emotions to predict the next word. EmoWrite employs a Recurrent Neural Network (RNN) for predicting contextualized words. RNN has been proven to be the most efficient machine-learning algorithm \cite{29} that provides consistent refinement in the system by requiring less feature engineering, which is a time-consuming task. It also effectively adopts new data and has parallel processing abilities. RNN has an advantage over other neural networks as all the inputs are dependent on each other, it keeps track of relations with previous words and helps in anticipating preferable output. Being an online algorithm, it updates itself after a specific period when new data is entered by the user.

\subsubsection{AutoComplete}
The autocomplete feature of the EmoWrite system predicts the word on a character-to-character basis. For this purpose, four different datasets are created, each containing words related to an emotional class. The words are predicted from the respective dataset as per the user’s emotional state.

\subsubsection{Communication Interface}
The communication interface comprises a GUI through which a user interacts with the system by selecting words or characters from the GUI through mapped brain signals. The dynamic arrangement of the virtual keyboard uses a character sequence pair to display the next set of characters depending on the last text entered. The RNN model is also used here to display characters according to the user’s typing style.

\subsubsection{Feedback}
Usually, feedback is given through a visual or auditory stimulus. EmoWrite uses visual feedback; the user gets feedback through previously written text which is then displayed on the top of the screen. This process provides continuous learning, by storing the latest written text in the database and then feeding it to the machine learning algorithm to update the model. Every time, the latest available model is used for prediction ensuring coherent predictions and helping the user in effective system manipulation and trouble-free embracing.

\subsubsection{Interface Arrangement}
To address the problems of GUI present in existing systems, a circular dynamic keyboard is designed to reduce traversing time, use screen space efficiently, and reduce the distance between characters \cite{11}. Only limited characters are shown on the virtual keyboard at a time, and the next set of characters to appear on the screen is dependent on the previously accessed character. The appearance of the next set of characters uses the character sequence pair model and machine learning algorithm, where the character sequence pair estimates the probability of occurrence of character pairs i.e., it estimates the occurrence of certain characters against previously typed characters. This probability is calculated using approximately 3.2 million characters from seven English-language novels \footnote{http://homepages.math.uic.edu/\~leon/mcs425\-s08/handouts/char\_freq2.pdf}. Initially, the most used characters appear on the screen. For example, the first set of characters that appears on the keys is shown in Figure \ref{fig4a}. After the selection of character ‘t’, the next set of characters appears on the screen, depending on the typed character ‘t’. The next character set is shown in Figure \ref{fig4b}.

\subsubsection{Emotiv Commands for Interface Navigation}
Control commands of the brain are mapped to control traversal in the interface. The commands used to control the interface are shown in Table \ref{tab2}. Some mental states, emotional classes, and facial expressions are mapped with basic functionalities to navigate the interface. Mental states are used to control the navigation direction, and facial expressions while motor imagery movement is used to transfer focus from one section to another.

\begin{table*}[ht!]
\centering
\small
\caption{List of Emotiv Commands for Interface Navigation}
\label{tab2}
\begin{tabular}{|c|c|c|} \hline
                               & Commands   & Actions                                       \\ \hline
\multirow{4}{*}{Mental State}  & Left       & Left movement         \\ \cline{2-3}
                               & Right      & Right movement        \\ \cline{2-3}
                               & Pull       & Up movement           \\ \cline{2-3}
                               & Push       & Down movement         \\ \hline
Facial Expression              & Smile      & Selection             \\ \hline
\multirow{2}{*}{Motor Imagery} & Look Right & Focus shifts toward the Helping Verb section \\ \cline{2-3}
                               & Look Left  & Focus shifts toward the Prediction section  \\  \hline
\end{tabular}
\end{table*}

\subsubsection{Conversion of Thought to Text}
To convert thought to text, the mental commands are detected through the brain signals $I_s$ and compared with the given trained signals $T_s=\{ left,\ right,\ up,\ down \}$. $I{_{s_i}}(t)$ is the input signal of the $i^{th}$ channel at time $t$ which will be compared with the trained signals.

\begin{equation}
    M\left(t\right)=\sum_{i=1}^{14}\frac{I_{s_i}(t)}{T_s}
    \label{eq1}
\end{equation}

So, the mental commands $M\left(t\right)$ at time t can be measured by dividing the extracted signal at time $t$ with the trained data. It can be illustrated in Equation \ref{eq1}, where $i$ is the number of channels. The detected command will be:

\begin{equation}
    D_c=M_\propto\left(t\right),\ \ \ \ \ \ \ \ \ \ \ \ \ \ \ \ \ \ \ \propto>0.80
    \label{eq2}
\end{equation}

Here, $\propto$ is the threshold level or the confidence level. It should be greater than 0.80 for accurate detection of brain signals. The desired command changes the focus of the keyboard. Initially, the focus will be on the center of the keyboard i.e., the space key but after the detection of a mental command from Equation \ref{eq2}, the focus will change accordingly. $B(D_c)$ in Equation \ref{eq3} is the button in the direction $D_c$.

\begin{equation}
    B(D_c)=\left\{\begin{matrix}1&D_c\in T_s\\0&D_c\notin T_s\\\end{matrix}\right.
    \label{eq3}
\end{equation}

The ``1'' value means the focused key has a yellow color. In this way, the user can change focus on any key of the keyboard. After changing the focus, the focused character can be selected using Equation \ref{eq4}, where $f$ is the user’s facial expression, chosen from the set $F_E\ = \{ Blink,\ Wink,\ Surprise,\ Frown,\ Smile,\ Clench,$ $\ Laugh,\ Smirk \}$ and $F_o$ is the frequency of occurrence defined as set $F_o= \{ once,\ twice,\ thrice \}$. 

\begin{equation}
    S_L\left(D_c\right)=\left\{\begin{matrix}1&f=\mathrm{Blink},f\in F_E\land F_o=twice\\0&f\neq B l i n k\\\end{matrix}\right.
    \label{eq4}
\end{equation}

Equation \ref{eq4} implies that the selection will occur only if the facial expression $F_E$ is a blink and it is done twice. The ``1'' value here means the selection of a character while ``0'' means no selection. The selected character/label $S_L$ will be written to the text area $T$.

$U =\{ e, \ t, \ a, \ o, \ i, \ n, \ s, \ r, \ h, \ l, \ d, \ c, \ u, \ m, \ f, \ p, $ $ \ g, \ w, \ y, \ b, \ v, \ k, \ x, \ j, \ q, \ z \}$ represents the set of alphabets placed according to their frequency of occurrence. so the circular keyboard contains the first 6 characters from the set $U$. The labels on the keys can be represented by a matrix ${Dis}_k$ having two rows and four columns where rows represent the number of circles, and the columns represent the number of keys in each circle.

\begin{equation} 
{Dis}_k=\left|\begin{matrix}\begin{matrix}e&t\\i&n\\\end{matrix}&\begin{matrix}a&o\\\gets&more\\\end{matrix}\\\end{matrix}\right|
\label{eq5}
\end{equation}

Initially, characters having the highest frequency will be displayed on the screen shown in Equation \ref{eq5}. After the selection of a specific character from Equation \ref{eq4}, the next set of characters that will appear on the screen will be dependent on the likelihood of occurrence of each alphabet after $S_L$ and is determined using Equation \ref{eq6} which calculates the probabilities of all alphabets to the typed ($x$). 

\begin{equation}
    PNextChar\left(x\right)=\prod_{i=1}^{26}\frac{P(x_i)}{P(x)} \ \ \ \   x\equiv S_L, {\ \ \ \ x}_i\in U
    \label{eq6}
\end{equation}

Suppose the selected label $S_L$ to be `$e$' then the next character probabilities will be $PNextChar\left(e\right) = \{ a\left(0.01\right),b\left(0.0023\right),.\ .\ .\ ,z(0.00) \}$ and the display matrix ${Dis}_e$ will be:

\begin{equation}
{Dis}_e=\left|\begin{matrix}\begin{matrix}r&d\\a&t\\\end{matrix}&\begin{matrix}s&n\\\gets&more\\\end{matrix}\\\end{matrix}\right| \nonumber
\end{equation}

Here, the first most probabilistic set of `$e$' will be displayed, and if the desired character is not present in the set of displayed characters, the user can click the key with the label `$more$'. The next probabilistic set will then be displayed. 

Given the singular and plural helping verb $Sin = \{ is, $ $ \ am,\ was, \ has,\ the \}$ and $Plu = \{ are, were, have, a, the \}$, the prediction of the helping verb $H_v$  will be dependent on the written text T. Suppose $S$ and $P$ are the set of singular and plural words respectively, then helping verb is predicted using Equation \ref{eq7}.

\begin{equation}
    H_v(T)=\left\{\begin{matrix}x&T\in S,\ x\in S i n\\y&T\in P,\ y\in P l u\\\end{matrix}\right.
    \label{eq7}
\end{equation}

\begin{figure*}[ht!]
\centering
\includegraphics[width=1.0\textwidth]{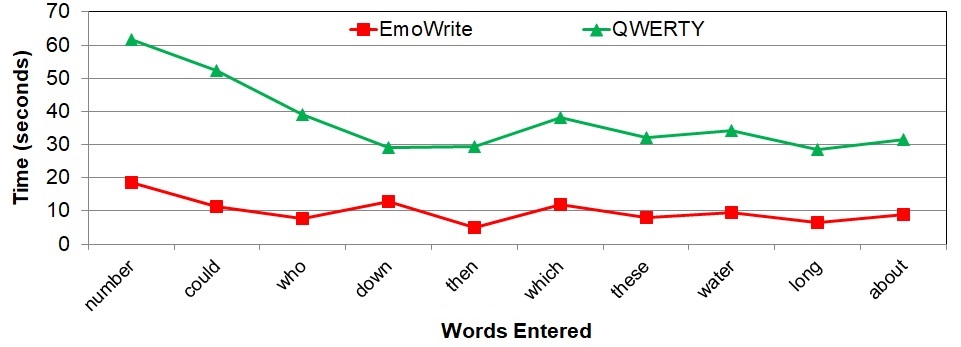}
\caption{Total time required to type the 10 words using QWERTY and EmoWrite keyboards}
\label{tspeed}
\end{figure*}

The word prediction will be dependent on the context and emotional state of the user. The emotional state of the user depends on the valence $E_V$ and arousal $E_A$. Valence is the negativity or positivity of emotion which can be measured by comparing hemispherical activation and arousal is the activation level of the brain.

\begin{eqnarray}
\centering
    \begin{matrix}E_A\left(t_f-t_i\right)\rightarrow h i g h,low\\E_V\left(t_f-t_i\right)\rightarrow p o s i t i v e, negative \\D_E=E_A+E_V\end{matrix}
    \label{eq8}
\end{eqnarray}

$t_i$ and $t_f$ are the initial and final time respectively, and the detected emotion $D_E$ can be one of the four categories $D_E = \{ happiness,\ sadness,\ anger,\ calm \}$. For the prediction of the next word, Recurrent Neural Network (RNN) is used. It is good at learning sequential and temporal data. It also learns the word-level features. The word prediction is based on previously written sentences of the n-word. Equation \ref{eq9} gives us the probability of observing a sentence.
\vspace{-1mm}
\begin{equation}
    P\left(m_1,.\ .\ .,m_n\right)=\prod_{j=1}^{n}{P(m_j|m_1,.\ .\ .,m_{j-1})}
    \label{eq9}
\end{equation}

First, the sequence of sentences will be converted into a sequence of words w where $w_i$ represents a single word. Each word will be represented as a set of elements equal to the vocabulary size $V_s$ and the sequence of words will become a matrix that will be given as an input to the RNN. Three parameters $X$, $Y$, and $Z$ are used which represent the input to a layer, output to a layer, and output towards the next state respectively. The equations of RNN are:

\begin{equation}
    s_t=\tanh(X_{w_t}+Z_{s_{t-1}})
    \label{eq10}
\end{equation}

\begin{equation}
    O_t=softmax(Y_{s_t})
    \label{eq11}
\end{equation}

Here, $s_t$ is the state at the time $t$ and $O_t$ is the output at time $t$. Using hidden layer $H = 100$, we have $w_t\in\ R^{8000}$, $O_t\in\ R^{8000}$, $s_t\in\ R^{100}$, $X\in\ R^{100\ast8000}$, $Y\in\ R^{8000\ast100}$, and $Z\in\ R^{100\ast100}$.

First, we apply forward propagation that will predict the word probabilities and return a state as output. Then we predict that results in the highest probability word. After predicting the word, we must calculate the loss, to check whether our prediction is correct or not. A loss should be minimal and can be calculated using equation \ref{eq12} that shows the loss concerning the prediction $O$ and true label $t$ on words in the text (training example) $W$. The greater the difference between the output and the true label, the greater the loss.

\begin{equation}
    L\left(t,O\right)=-\frac{1}{W}\sum_{n\in W}{t_nlogO_n}
    \label{eq12}
\end{equation}

Let $RNN_W$ be the words predicted by RNN and $H$, $S$, $A$, and $C$ be the sets that include words from the $happiness$, $sadness$, $anger$, and $calm$ classes respectively. Then the emotion-based word prediction using Equations \ref{eq8} -\ref{eq11} is:

\begin{equation}
    \footnotesize{EmoPred(D_E,O_t)}=\left\{\begin{matrix}\begin{matrix}p&D_E=happy,p\in(H\cap{\rm RNN}_W)\\q&D_E=sad,p\in(S\cap{\rm RNN}_W)\\\end{matrix}\\\begin{matrix}r&D_E=angry,p\in(A\cap{\rm RNN}_W)\\s&D_E=calm,p\in(C\cap{\rm RNN}_W)\\\end{matrix}\\\end{matrix}\right.
    \label{eq13}
\end{equation}

\section{Results}
\label{ER}
To prove the efficiency and productivity of EmoWrite, the system is evaluated based on parameters, accuracy, ease of use, and words per minute count. To check these parameters, the following experiments are performed.

\begin{figure*}[ht!]
\centering
\includegraphics[width=1.0\textwidth]{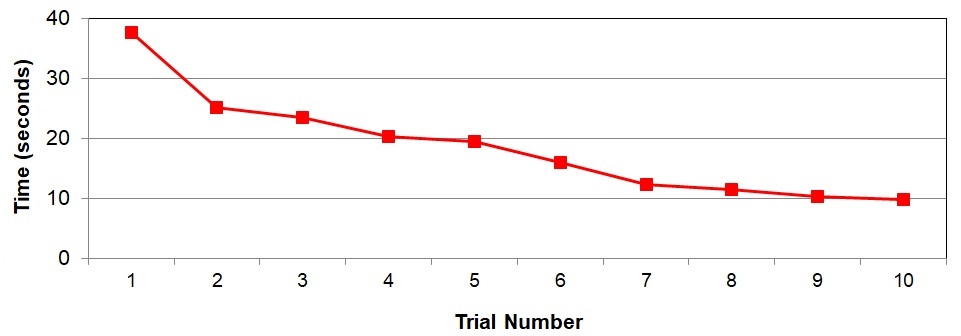}
\caption{Average time required to type a word over 10 trials using EmoWrite keyboard}
\label{efficency}
\end{figure*}

\subsection{Keyboard Efficiency}
To evaluate the efficiency of the dynamic keyboard, we use two keyboards in the experiment, i.e.,  (1) a QWERTY keyboard with scan-through keys, and (2) a Dynamic keyboard (EmoWrite). There are two rounds in this experiment. In the first round, the participants are asked to write 10 words (i.e., number, could, who, down, then, which, these, water, long, and about) using each keyboard, one by one. The starting time, when the user starts thinking of the command, and the ending time, when the participants are done with writing the word, is noted and the total time taken is calculated for each word. While in the second round, the participants are asked to write a whole sentence, in this case ``the Brain-computer interface helps in processing brain signals'', using each keyboard, and its total time is also calculated. Figure \ref{tspeed} shows the total time required to type these words using both keyboards. It is observed that EmoWrite takes less time to type in all the words while the QWERTY keyboard with scan-through keys requires more time to type in all the words.  

We also determined the time taken to type a complete sentence using the QWERTY and EmoWrite keyboards. It took the participants, on average, 7 minutes and 5 seconds to type a complete sentence using the QWERTY keyboard while it only took 2 minutes and 20 seconds to type the same sentence using EnoWrite. 

Moreover, another experiment is performed to check the efficiency of the system that comprises 10 trials. In each trial, the participants are asked to write 10 words of 3 to 8-character length with the help of brain signals. The time to type each word and time per character is recorded. Time per character is calculated by dividing the time to type each word by the total number of characters. Then, the average time to write a word and a character is calculated. Figure \ref{efficency} shows the average time to type a word over 10 trials. It is observed that the average time to type a word gradually decreases with each trial as the participants become more familiar and trained with the system in generating specific brain signals efficiently.

\subsection{Words per Minute}
To compare the performance of EmoWrite with existing approaches, we calculate the total number of words typed per minute. For this experiment, the participants are asked to write the following given sentences with the help of brain signals; i.e. ``f you watch the hills in London you will realize what torture it is”, ``it is so annoying when she starts typing on her computer in the middle of the night”, ``sucks not being able to take days off from work”, ``her hotel is restricting how the accounts are done adds a bit more pressure”, ``I was thinking about how excited I am for you guys to move”. These sentences comprise 68, 83, 47, 77, and 57 characters respectively. The participants were timed for one minute. After the completion of one minute, the participants are asked to stop writing and a total number of words and characters per minute are recorded. Table \ref{comparison} shows the comparison of the CPM and WPM of EmoWrite with previously proposed systems. \cite{10} and \cite{7} have implemented the next word prediction feature, while \cite{12} implements the next character prediction feature and the remaining systems do not have any predicting features integrated. EmoWrite outperforms existing systems with the highest CPM and WPM.

\begin{table}[ht!]
\centering
\caption{Comparison of CPM and WPM of EmoWrite with existing studies}
\label{comparison}
\begin{tabular}{|l|c|c|} \hline
Reference     &  CPM  & WPM \\ \hline
Zhang et al.\cite{2}  		& 6.7  & 1   \\ \hline
Alomari et al.\cite{6} 		& 7.0  & 1   \\ \hline
Pandarinath et al.\cite{10} & 6.6  & 1   \\ \hline
Hayet et al.\cite{7} & 12.0 & 2   \\ \hline
Pathirana et al.\cite{12} 	& 25.0 & 5.1 \\ \hline
EmoWrite 					& 31.9 & 6.6 \\  \hline
\end{tabular}
\end{table}


\subsection{Information Transfer Rate}
We calculate the information transfer rate (ITR) for each word and each command which is the total time taken over the total number of actions performed. ITR for commands and letters is calculated using Equations \ref{eq14}-\ref{eq15} where $N_c$ is the total number of possible commands, $C_N$ is the number of commands required to write an N letter word, $N_l$ is the total number of characters in the keyboard, $L_N$ is the total number of letters in a word, and $t$ is the total time.

\begin{equation}
    {ITR}_c={log}_2\left(N_c\right).C_N/t
    \label{eq14}
\end{equation}
\vspace{-5mm}
\begin{equation}
    {ITR}_l={log}_2\left(N_l\right).L_N/t
    \label{eq15}
\end{equation}

Table \ref{itr} shows the ITR for commands and letters. The average information transfer rate of commands is 87.55 bits/min and of letters is 72.52 bits/min. These information transfer rates are more than the rates of \cite{22}, where the information transfer rate for commands is 62.71 bits/min, and for letters, it is 50.14 bits/min.

\begin{table}[ht!]
\centering
\caption{Comparison of ITR in bits/minutes of EmoWrite with existing study}
\label{itr}
\begin{tabular}{|c|c|c|} \hline
Ref.     & Commands & Letters \\ \hline
Cecotti\cite{22} & 62.71    & 50.14   \\ \hline
EmoWrite & 87.55    & 72.52   \\ \hline
\end{tabular}
\end{table}

\subsection{Accuracy and Latency}
To evaluate the accuracy of the EmoWrite system, we determine the difference between the intended target and the observed target. The intended target is the character the participants aim to select while the observed target is the character selected based on brain signals, potentially due to errors. In the experiment, participants are instructed to speak aloud about what they intend to write, and the system records the selected character. Each written character is then marked as either correct or incorrect. The accuracy is computed by calculating the total number of correctly written characters compared to the total number of characters attempted. The resulting accuracy of the proposed EmoWrite system is 90.36\%.

Furthermore, to check the latency of EmoWrite, we measure the mean and standard deviation of delay (which is the time taken by the user when he/she starts thinking of a command until he/she performs it). The user is shown a certain command (left, right, up, or down), and the time is noted until the user successfully achieves that target. The mean delay of EmoWrite is 2.685 seconds, which is less than the delay of the system proposed by \cite{30} which has a mean delay of 3 seconds.

\subsection{Participant Stress Monitoring during System Interaction}
To assess participant responses to the system, we monitored stress levels through facial expressions. If a participant exhibited signs of stress, such as frowning, it indicated potential difficulty with the system. When stress reached a predefined threshold, a notification was displayed. Analysis revealed that fewer than 20\% of measurements indicated signs of stress, suggesting the system is generally well-received and does not induce significant stress. This approach provides valuable insights into participant reactions. However, we acknowledge that additional methods could further enhance our understanding of user experience.

\subsection{Effectiveness of Integrating Emotion-based Prediction}
\begin{figure}[ht!]
\centering
\includegraphics[width=0.47\textwidth]{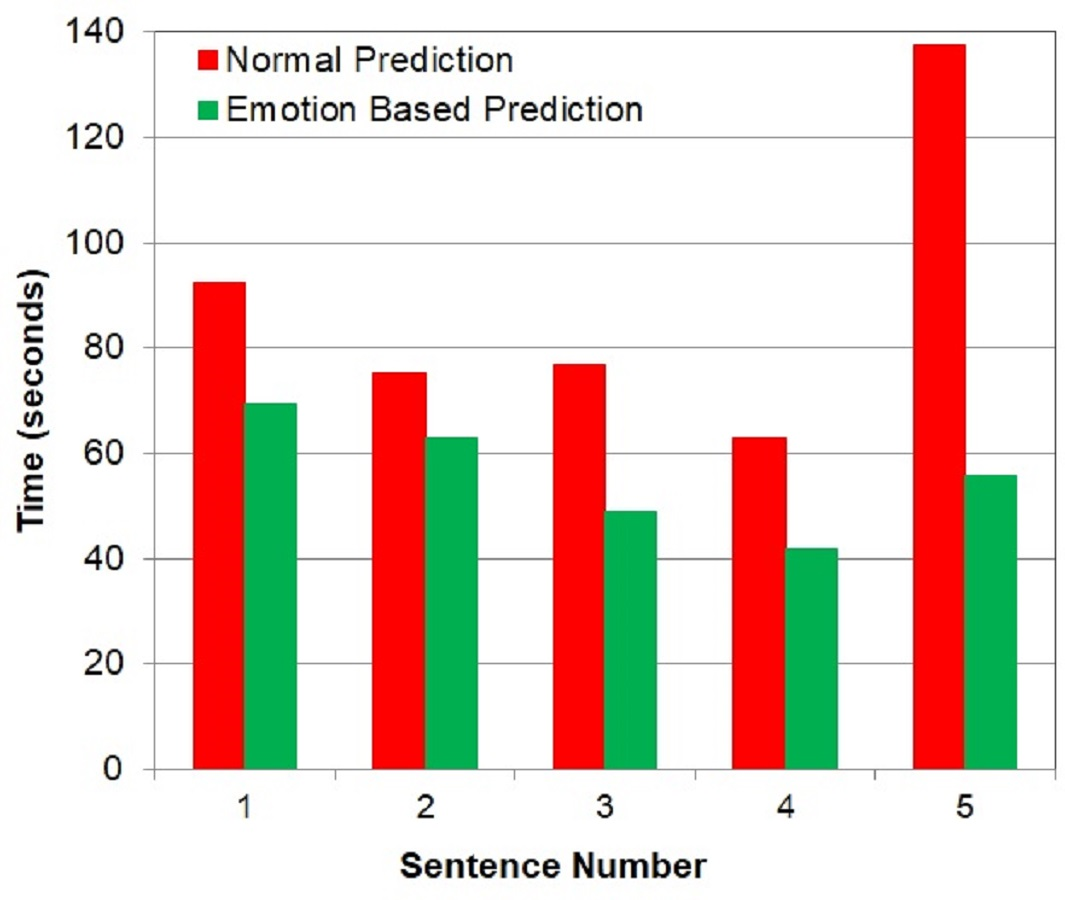}
\caption{Average time required to type sentences with and without incorporating emotional states.}
\label{empre}
\end{figure}

The novelty of this system is the emotion-based predictions, and to check its effectiveness, we conduct an experiment consisting of two rounds. In the first round, the participants are asked to write some simple sentences without observing their emotional state. The sentences used in the first round are ``I am tired of my job”, ``I don’t like this world”, ``That was awesome”, ``I love this world”, and ``I have an infection”. In the second round, the participants are asked to write these sentences again but with the integration of their emotional states. For this, the participants are first shown some video depicting an emotional state. The difference is observed in the word predictions concerning the emotional states i.e., happiness, anger, sadness, and calm. Figure \ref{empre} shows the difference in the average time required to type the sentences with and without emotional states. It is observed that using emotion-based prediction is more effective as it gives a prediction of emotion-related words like ``horribly'', ``ughhh'', ``awesome'' and ``terrible'', and this requires less time to type.

\section{Discussion}
\label{DIS}
This paper introduces EmoWrite, a system designed to aid paralytic individuals in converting their thoughts into text using a sentiment analysis approach. EmoWrite leverages EEG signals, facial expressions, and emotional states to facilitate effective communication.
The system has demonstrated promising results in terms of accuracy and typing speed. However, the study acknowledges certain limitations and areas for improvement:

\subsection{Training Time}
One of the primary limitations is the extensive training time required for participants to effectively use EmoWrite. Each user must undergo a substantial training period to achieve a high level of proficiency with the system. This training involves learning to control the interface with mental commands and facial expressions, as well as adapting to the personalized predictive text system.

\subsection{Failure Cases}
The analysis of failure cases in this study is based on the accuracy of the results. Instances where the system makes incorrect predictions are evaluated to understand the underlying causes and improve the system's robustness. These errors can be attributed to factors such as signal noise, user fatigue, and variations in individual brain signal patterns.

\subsection{Generalizability}
While EmoWrite has been tailored to suit individual users by incorporating their emotional states and typing patterns, its generalizability to a broader population with varying types of paralysis and cognitive conditions remains an area for further research.

\subsection{User Interface}
The dynamic and adaptive nature of the virtual keyboard in EmoWrite is a significant advancement, but the interface's complexity can pose a challenge to new users. Simplifying the interface without compromising functionality could enhance user experience and reduce the learning curve.

\subsection{Future Improvements}
Future iterations of EmoWrite could benefit from several enhancements to further its effectiveness. Incorporating advanced machine learning algorithms could improve the system's accuracy and reduce prediction errors. Additionally, exploring alternative EEG headsets with higher signal resolution and reduced noise may enhance the system's overall performance and reliability.

Moreover, a direct measure of cognitive load should be integrated into the system's evaluation to provide a more comprehensive understanding of usability. Improvements in user training protocols and refining the system’s accuracy and usability will also be key focus areas.

In conclusion, EmoWrite represents a significant advancement in assistive communication technologies for individuals with severe disabilities. While the current version demonstrates innovative potential through its integration of sentiment analysis with thought-to-text conversion, addressing its limitations through ongoing research and development will be crucial in maximizing its impact and enhancing the quality of life for users.


\section{Conclusion}
\label{con}
In summary, we have presented a pioneering approach to converting silent speech into text, revolutionizing interaction for individuals with paralysis. Our proposed solution leverages brain signals to establish a controlled interface, empowering paralytic patients to engage with the world. Key components of this interface include a dynamic circular keyboard, word prediction, and a segment dedicated to aiding verbs. Recognizing the critical impact of keyboard design on typing speed, we employ a circular layout to minimize traversal delays. The arrangement of characters dynamically displays a limited set on the screen, curtailing delay and enhancing typing speed. Integrating machine learning algorithms, we capture user writing patterns, facilitating predictive words, and aiding verb suggestions. Notably, emotion-driven predictions further streamline the user experience, enabling the auto-completion of entire words based on context. Our approach's efficacy was verified through tests involving novice users, assessing parameters like Words Per Minute (WPM), user-friendliness, and system accuracy. The system's standout features encompass the dynamic character arrangement, emotion-enhanced word predictions, and user-specific contextual character display. EmoWrite yields an impressive 90.36\% accuracy in thought-to-text conversion, achieving 6.58 words and 31.92 characters per minute. Importantly, Information Transfer Rates (ITR) for commands and letters stand at 87.55 and 72.52 bits/min respectively, accompanied by a latency of 2.685 seconds. These results collectively underscore the potency of our innovative system, setting new benchmarks in enhancing communication efficiency and usability for individuals with motor disabilities.

\section*{Conflicting interests}
"The authors have no conflict of interests."


\section*{Ethics Declarations}
``Prior to study initiation, written informed consent was obtained from all participants and the study protocol was approved by the Ethics Committee of COMSATS University Islamabad, Lahore Campus. All experiments were performed in accordance with relevant guidelines and regulations. Informed consent was obtained from all participants for publication of identifying information/images in an online open-access publication.''

\section*{Guarantor}
"Not applicable."

\section*{Contributorship}
"Not applicable."


\end{document}